\documentclass[conference]{IEEEtran}
\IEEEoverridecommandlockouts
\usepackage{marvosym}

\usepackage{amsmath,amssymb,amsfonts}
\usepackage{algorithmic}
\usepackage{textcomp}
\usepackage{xcolor}
\usepackage{subfigure}
\usepackage{hyperref}
\usepackage{cleveref}
\usepackage{colortbl}
\usepackage{tikz}
\usepackage{graphicx} 
\usepackage{pgfplots}
\pgfplotsset{compat=1.14}       
\usepackage{booktabs}
\usepackage{multirow}
\usepackage{makecell}
\usepackage{tabularx}
\usepackage{array}
\usepackage{xtab}
\usepackage{longtable}
\usepackage[utf8]{inputenc}
\usepackage{url}
\usepackage{cite}
\usepackage{float}
\usepackage{subcaption}
\usepackage{makecell}
\usepackage{filecontents}
\usepackage[normalem]{ulem} 
\def\BibTeX{{\rm B\kern-.05em{\sc i\kern-.025em b}\kern-.08em
    T\kern-.1667em\lower.7ex\hbox{E}\kern-.125emX}}

\begin{document}

\title{Agents4PLC: Automating Closed-loop PLC Code Generation and Verification in Industrial Control Systems using LLM-based Agents\\
}

\author{\IEEEauthorblockN{
Zihan Liu*}
\IEEEauthorblockA{
\textit{Zhejiang University}\\
zihanliu@zju.edu.cn
}
\and
\IEEEauthorblockN{
Ruinan Zeng*}
\IEEEauthorblockA{
\textit{Zhejiang University}\\
zengruinan@zju.edu.cn
}
\and
\IEEEauthorblockN{Dongxia Wang\textsuperscript{\Letter}\thanks{\noindent Co-first authors: Zihan Liu and Ruinan Zeng; Corresponding authors: Dongxia Wang and Wenhai Wang}}
\IEEEauthorblockA{
\textit{Zhejiang University}\\
dxwang@zju.edu.cn
}
\and
\IEEEauthorblockN{Gengyun Peng}
\IEEEauthorblockA{
\textit{Zhejiang University}\\
pgengyun@zju.edu.cn
}
\and
\IEEEauthorblockN{Jingyi Wang}
\IEEEauthorblockA{
\textit{Zhejiang University}\\
wangjyee@zju.edu.cn
}

\and
\IEEEauthorblockN{
Qiang Liu}
\IEEEauthorblockA{
\textit{Zhejiang University}\\
22460396@zju.edu.cn
}
\and
\IEEEauthorblockN{Peiyu Liu}
\IEEEauthorblockA{
\textit{Zhejiang University}\\
liupeiyu@zju.edu.cn
}
\and
\IEEEauthorblockN{Wenhai Wang\textsuperscript{\Letter}}
\IEEEauthorblockA{
\textit{UWin Tech \& Zhejiang University}\\
zdzzlab@zju.edu.cn
}
}


\maketitle

\begin{abstract}
In industrial control systems, the generation and verification of Programmable Logic Controller (PLC) code are critical for ensuring operational efficiency and safety. While Large Language Models (LLMs) have made strides in automated code generation, they often fall short in providing correctness guarantees and specialized support for PLC programming. To address these challenges, this paper introduces Agents4PLC, a novel framework that not only automates PLC code generation but also includes code-level verification through an LLM-based multi-agent system. We first establish a comprehensive benchmark for \emph{verifiable PLC code generation} area, transitioning from natural language requirements to human-written-verified formal specifications and reference PLC code. We further enhance our `agents' specifically for industrial control systems by incorporating Retrieval-Augmented Generation (RAG), advanced prompt engineering techniques, and Chain-of-Thought strategies. Evaluation against the benchmark demonstrates that Agents4PLC significantly outperforms previous methods, achieving superior results across a series of increasingly rigorous metrics. This research not only addresses the critical challenges in PLC programming but also highlights the potential of our framework to generate verifiable code applicable to real-world industrial applications.

\end{abstract}

\begin{IEEEkeywords}
Code Generation, Code Validation, PLC Code, LLM-based Agents, Multi Agents, Industrial Control System
\end{IEEEkeywords}
\section{Introduction}


Programmable Logic Controllers (PLCs) are essential components of Industrial Control Systems (ICSs), playing a crucial role in industrial automation and management of key industrial processes. 
The global PLC market is projected to reach USD 12.20 billion by 2024, with a compound annual growth rate (CAGR) of 4.37\% from 2024 to 2029 \cite{mordor_plc_market_2024, Technavio_plc_market_2023}. 
This growth is largely driven by the increasing reliance on 
industrial control programming languages based on IEC 61131-3 standard \cite{IEC61131-3}, such as Structured Text (ST) and Function Block Diagram (FBD), to oversee and regulate critical infrastructure systems across key sectors like energy \cite{wang2021application}, manufacturing \cite{schreyer2000design}, and transportation \cite{kornaszewski2020use}. Among others, ST language, as a text-based language, is most similar to other popular high-level languages (in terms of syntax and program structure) and is thus suitable for code generation. 

In ICSs, automatic generation of control code can greatly reduce repetitive tasks, significantly enhancing engineers' productivity. 
With the rapid advancement of large language models (LLMs), automatic code generation has gained much attention across various programming languages (e.g., C \cite{xu2024automated}, C++ \cite{chen2024supersonic}, Python \cite{zhang2024codeagent}, Java \cite{corso2024generating}, etc) due to its potential to automate software development and reduce costs. It is thus desirable to explore LLM-based automated code generation methods for industrial control code development.


Emerging LLMs for code generation (a.k.a. code LLMs), such as Openai Codex \cite{chen2021evaluating}, AlphaCode \cite{li2022competition}, CodeLlama \cite{roziere2023code},
may not be ideal
for PLC code generation due to the following critical reasons.
Firstly, these code LLMs may excel at generating code in mainstream high-level languages such as C or Python, but mostly perform poorly for industrial control code. Indeed, it is notoriously challenging to collect sufficient data for fine-tuning a specialized model for industrial code due to the proprietary and specialized nature of PLC code. 
Secondly, as the control code is used to manage the operation of industrial sectors, it is of vital importance to guarantee their functional correctness, which is far more challenging than generating executable code.      
There exist some efforts towards achieving specialized PLC code generation, such as LLM4PLC \cite{fakih2024llm4plc} and the work of Koziolek \cite{koziolek2024llm-retrieval} either by finetuning or Retrieval-Augmented Generation (RAG) enhancements.
LLM4PLC \cite{fakih2024llm4plc} also incorporates workflows for syntax and functional verification at the design-level beyond code generation to improve the code quality.
However, \emph{the correctness of specifications is only verified at the design level while the correctness of the LLM-generated code remains questionable.} Moreover, their model-centred architecture lacks agility in integrating the whole development pipeline to achieve full automation.
On the other hand, there is a growing trend in automated LLM-based development workflows aimed to further refine or validate code generated by LLMs using a multi-agent system architecture. For instance, ChatDev \cite{qian2024chatdev} and MapCoder \cite{islam2024mapcoder} implement software development systems composed of multiple intelligent agents, each with distinct roles and tasks, with the goal of generating high-quality code in a closed-loop manner. Notably, such an agent-based framework is flexible to implement multiple relevant software engineering tasks like validation and debugging, which are crucial in improving the quality of generated code. 

In this work, we present a novel LLM-based multi-agent framework designed to address the limitations of current PLC code generation solutions, namely Agents4PLC. Our system comprises multiple agents, each tailored to specific tasks such as PLC code generation, syntax validation, functional verification and debugging (in case of failures). 
Such a closed-loop workflow allows Agents4PLC to effectively coordinate different agents to automatically generate high-quality code (verifiable correct in the ideal case). Note that different from LLM4PLC which only verified the correctness of the specification, we directly verify the correctness of the generated code.    
Meanwhile, by leveraging advanced multi-agent architectures like LangGraph and MetaGPT, Agents4PLC is highly adaptable to a wide range of PLC code generation tasks and incorporate different base code LLMs.


\begin{itemize} 

\item We establish a comprehensive benchmark that transitions from natural language requirements to formal specifications, utilizing verified reference code with human-checked labels, to facilitate future research in the field of PLC code generation.

\item We introduce Agents4PLC, the first LLM-based multi-agent system for fully automatic PLC code generation that surpasses purely LLM-based approaches (e.g., LLM4PLC \cite{fakih2024llm4plc}) by emphasizing code-level over design-level verification, and flexibility to incorporate various base code generation models (in both black-box and white-box settings) and supporting an array of tools for compilation, testing, verification and debugging.

\item We enhance our agents specifically for PLC code generation by implementing techniques such as Retrieval-Augmented Generation (RAG), advanced prompt engineering, and Chain-of-Thought methodologies, improving their adaptability and effectiveness in generating reliable PLC code.

\item We rigorously evaluate Agents4PLC against the largest benchmark available, demonstrating superior performance across a series of increasingly stringent metrics. We also deploy and validate our generated code on multiple practical scenarios, highlighting its potential for generating verifiable PLC code that meets the demands of practical industrial control systems.

\end{itemize}

\section{Background}
\subsection{PLC Programming Language}
PLCs are computer systems specifically designed for industrial automation control, enabling real-time monitoring and control of mechanical equipment and production processes. PLCs are widely utilized in manufacturing, transportation, energy, and other sectors due to their high reliability, ease of programming, and scalability. The IEC 61131-3 standard \cite{IEC61131-3} specifies five standard programming languages for PLCs, which include three graphical languages: Ladder Diagram (LAD), Function Block Diagram (FBD), and Sequential Function Chart (SFC), as well as two textual languages: Structured Text (ST) and Instruction List (IL). The structured text (ST) language, similar in syntax and structure to traditional programming languages such as C and Pascal, offers significant flexibility and readability. It supports common programming structures, including loops and conditional statements, making it widely used in scenarios that require complex mathematical calculations, data processing, and advanced control algorithms. With the development of Industry 4.0 and intelligent manufacturing, the prospects for ST language are extremely promising, as it can complement other programming languages to enhance the flexibility and efficiency of PLC systems.

\subsection{Syntax Checking and Functional Verification}
Code generated by LLMs often exhibits considerable uncertainty and may sometimes fail to compile or meet specified requirements \cite{siddiq2023lightweight}. Syntax checking ensures that the generated ST code is executable,
while functional verification ensures that the code realizes the expected functionality and avoids potential logic flaws or vulnerabilities \cite{first2022diversity}. 

\subsubsection{Syntax Checking}
A piece of code first needs to conform to the standards of a programming language before it can be compiled into an executable program. Many PLC Integrated Development Environments (IDEs), such as CODESYS and TwinCAT, can perform syntax checking for ST code. These IDEs conform to the IEC 61131-3 standard and provide functionalities such as programming, debugging, and simulation. However, due to the limitations of their platforms, these IDEs are not convenient for direct integration into automated code generation pipelines. Some command-line tools can also perform syntax checking and compilation of ST code. MATIEC \cite{MATIEC} is an open-source compiler capable of refactoring or compiling ST code into C language, widely used in the design and maintenance of industrial automation systems. RuSTy \cite{RuSTy} is an open-source project based on LLVM and Rust, aimed at creating a fast, modern, and open-source industry-grade ST compiler for a wide range of platforms, providing comprehensive compilation feedback.

\subsubsection{Functional Verification}
Functional verification is crucial for PLC code, ensuring that the generated ST code can accurately and reliably implement the intended control logic and operations in real-world applications, thereby preventing equipment failures or production interruptions due to logical errors or unforeseen circumstances. 
There exist some formal verification tools such as nuXmv \cite{cavada2014nuxmv} and PLCverif \cite{darvas2015PLCverif} for functional validation of ST code, which are serving as the backend verifier of our validation agent.

\subsection{LLM-based Agents}
The rapid advancement of LLMs has elevated traditional AI agents, which relied on reinforcement learning and symbolic logic \cite{xi2023rise, ribeiro2002reinforcement, kaelbling1996reinforcement, minsky1961steps, isbell2001social, liu2024large}, to a new level, i.e., LLM-based agents. With the assistance of LLMs, these agents have a strong ability to understand language, generate content, and utilize external knowledge and tools. They can perform complex tasks and make decisions through collaboration among agents and interaction with their environments.

\begin{figure}[!t]
\centering
\includegraphics[width=3.5in]{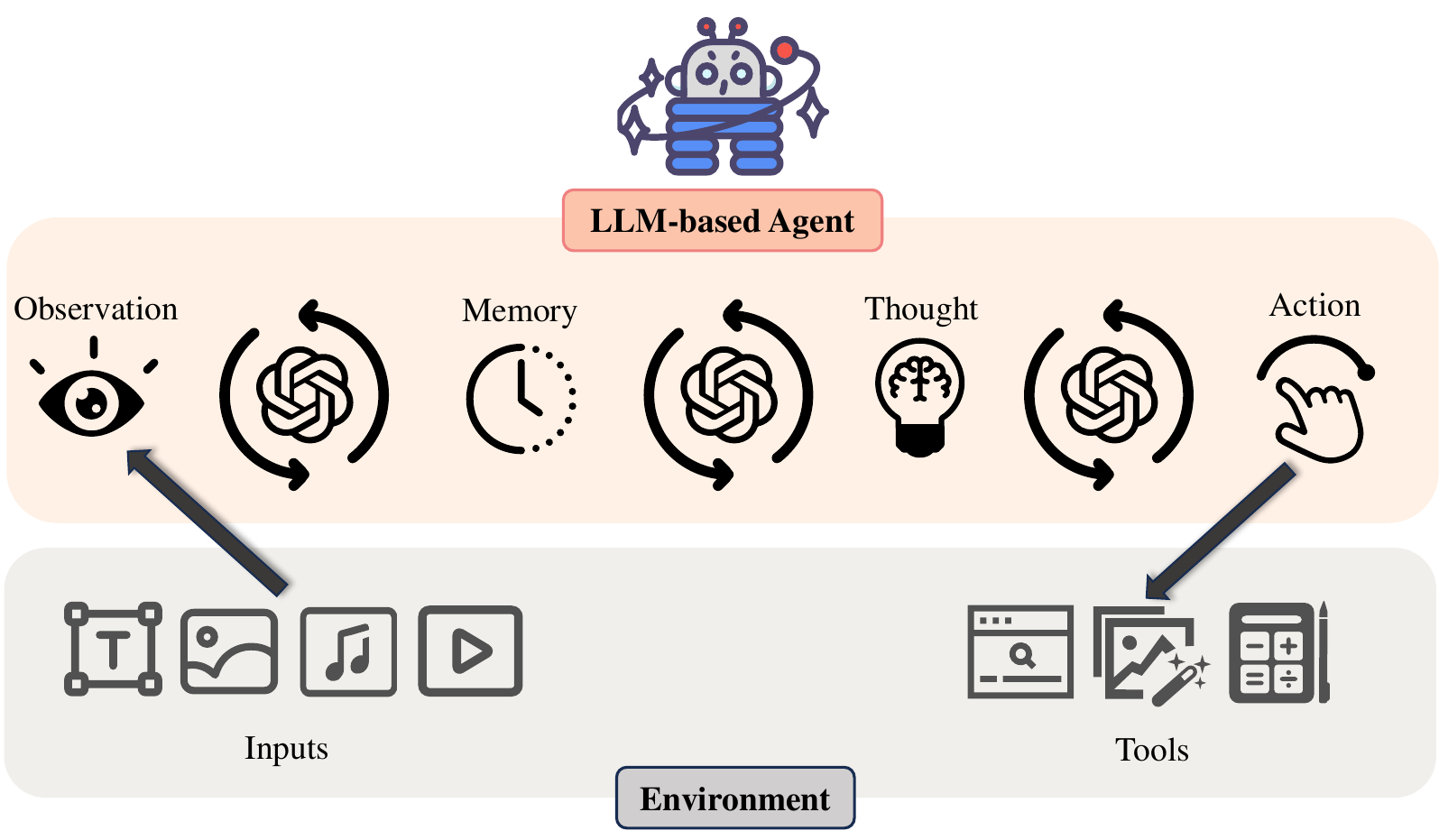}
\caption{Basic portrait of LLM-based agents}
\label{LLM-based agents}
\end{figure}
There exist various definitions of what comprises an LLM-based agent, and we present some common components here.
Typically, an LLM-based agent consists of five key components: LLM, observation, memory, thought and action \cite{hong2023metagpt}. Figure \ref{LLM-based agents} presents its basic framework\footnote{More details about different definitions and the components can be found in\cite{hong2023metagpt, zhang2024survey, wang2024survey}}. 
Each component can be analogized to human cognition for better understanding.
LLMs serve as part of an agent's “brain”, enabling it to comprehend information, learn from interactions, make decisions, and perform actions.
It perceives the environment with the observation component, e.g., receiving multimodal information from the others.
It uses memory component to store past interactions and observations.
It retrieves and analyzes information with thought component, to infer the next action.
And it execute actions using the LLMs or some external tools.


There are mainstream agent frameworks that can support the development of LLM-based multi-agent workflows, such as MetaGPT \cite{hong2023metagpt}, LangGraph \cite{LangGraph}, and AutoGen \cite{wu2023autogen}. These frameworks provide libraries that enable developers to create and manage multiple agents seamlessly. They facilitate the integration of large language models into complex workflows, allowing for efficient communication and collaboration among agents. Additionally, these frameworks often come with features like natural language processing capabilities, task orchestration, and easy scalability, making them ideal choices for building sophisticated multi-agent systems.

\subsection{Agent-based Software Development}
Agent-based software development is an emerging field that integrates autonomous and communicative agents to enhance various aspects of the software development life cycle. Recent advancements in this field such as ChatDev\cite{qian2024chatdev}, MetaGPT\cite{hong2023metagpt} and CodeAgent\cite{tang2024collaborative}, showcase the potential of integrating AI-driven agents to enhance collaborative software development frameworks, automate code review processes, and support contextual conversation within software development environments. These studies indicate a shift towards more interactive and intelligent software development tools that can adapt to the dynamic nature of development projects, providing developers with context-aware assistance and streamlining tasks through autonomous agents. The integration of advanced human processes within multi-agent systems emphasizes the potential of agent-based software development to revolutionize traditional software engineering practices.
\section{Agents4PLC Methodology}
\begin{figure*}[!t]
\centering
\includegraphics[width=7in]{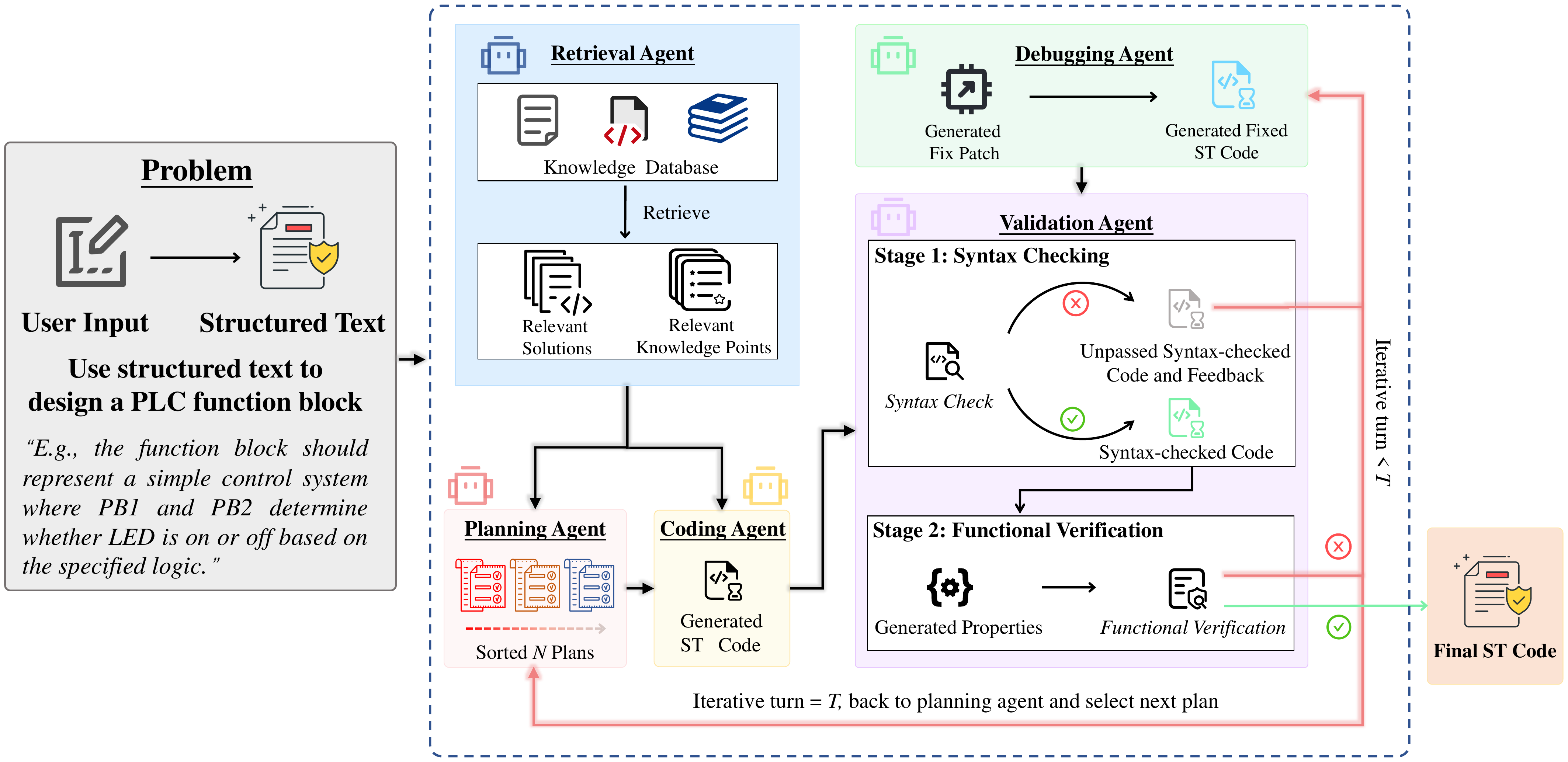}
\DeclareGraphicsExtensions.
\captionsetup{justification=centering}
\caption{Overview of Agents4PLC.}
\vspace{-3mm}
\label{fig:overview}
\end{figure*}

We propose Agents4PLC, an LLM-based multi-agent framework for automating the generation and verification process of ST code. The framework includes a set of agents with our defined roles and division of work, which automatically cooperate based on our defined workflow for the generation of reliable ST code. The whole framework is presented in Figure~\ref{fig:overview}. Below we present the detailed design of Agents4PLC.  

\begin{figure}[!t]
\centering
\includegraphics[width=3.6in]{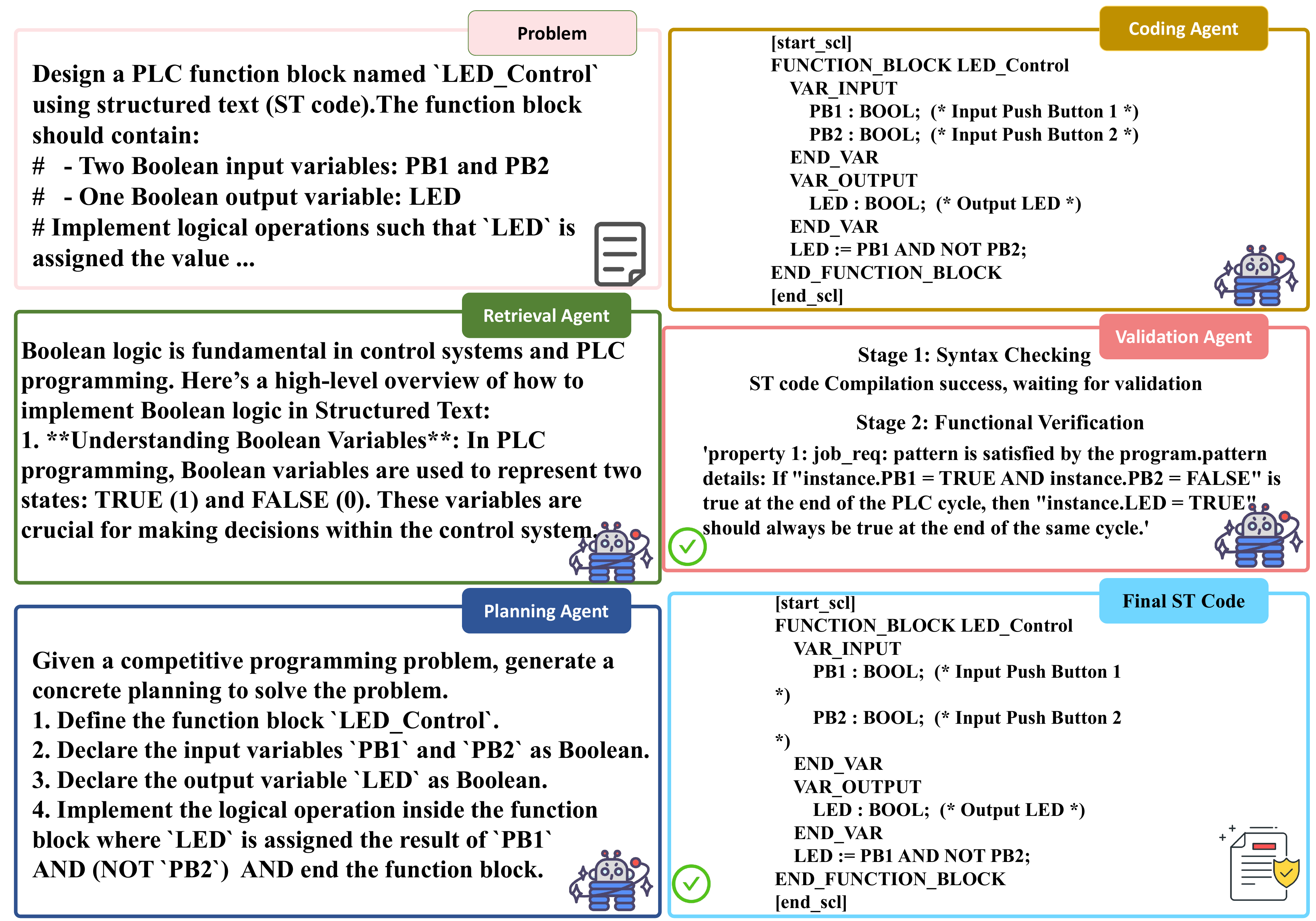}

\DeclareGraphicsExtensions.
\captionsetup{justification=centering}
\caption{An example of generating LED Control ST code with Agents4PLC.}
\vspace{-6mm}
\label{fig:example-overview}
\end{figure}

\subsection{Agents4PLC Framework}

As shown in Fig. \ref{fig:example-overview}, Agents4PLC takes an user input of ST coding requirement, e.g., to control an LED, and outputs the formally verified ST code. 
In this process, multiple agents with different roles cooperate to complete this task based on our carefully designed workflow. 
First, the retrieval agent analyzes the user input, based on which it retrieves the relevant information about ST codes, such as books and documents. Then it sends the retrieved information alongside with the user input (task instructions) to the planning agent, which is responsible to generate and rank the actionable plans for code generation, subsequently sent to the Coding Agent. The Coding Agent generates ST code accordingly to the received plans, and later a compiler checks whether there are syntax errors. If no error occurs, the validation agent (the verifier) will verify the functional correctness of the generated code. Otherwise, the error information will be delivered to the debugging agent who will return the fixing advice to the Coding Agent. If validation succeeds, the generation process is considered completed. Otherwise, the above workflow will iterate until it exceeds a predefined loop threshold, where the planning agent provides another plan for code generation.  

\subsection{Detailed Agents Design and Optimization}
The primary components of Agent4PLC are a group of autonomous agents powered by Large Language Models (LLMs) or analysis tools. 
These agents are meticulously designed to perform specialized tasks that collectively contribute to the efficient generation, validation, and iterative refinement of ST code. Each agent operates within a well-defined functional scope, allowing for the delegation of specific responsibilities such as code generation, debugging and validation.
In particular, the LLM-based agents excel at tasks such as understanding user specifications provided in natural language, retrieving relevant information, and generating ST code. 
Meanwhile, tool-based agents are designed to address domain-specific tasks, such as syntax checking and verification of ST code. The cooperation workflow among agents within this architecture not only enhances the scalability and adaptability but also facilitates collaborative interactions, creating a continuous feedback loop that improves the overall code development.


\subsubsection{Retrieval Agent} 
The Retrieval Agent is responsible for gathering relevant information for reference based on user input. Although some key agents in the framework are equipped with a RAG module to enhance task-specific abilities, the Retrieval Agent primarily focuses on searching for the relevant industrial control documents, PLC programming references, and ST code from a vector database for the reference of the following planning process. It can access both web-based search tools and also internal databases to extract pertinent information.


\subsubsection{Planning Agent} The Planning Agent receives the user input and the information retrieved by the retrieval agent, based on 
which it generates actionable plans using a structured automata-like format. Each plan represents a sequence of steps required to achieve the task. The planning agent ranks these plans, which later will get executed sequentially, based on their ranking, help identify the most suitable solution.

\subsubsection{Coding Agent}

The Coding Agent is pivotal in converting the detailed plans from the upstream agent into ST code, ensuring compliance with both syntactic and semantic requirements. By utilizing a RAG block, the agent accesses a comprehensive database of PLC documentation and validated ST code samples. RAG enables the agent to learn from the established experience and domain-specific knowledge, thus making informed decisions based on the provided PLC resources, help improve the reliability of the generated code.

In addition, prompts for the Coding Agent play a crucial role in steering the Coding Agent throughout the code generation process. These prompts encapsulate domain-specific rules and constraints, to guide the Coding Agent to follow the compiler requirements and the predefined validation criteria. The Coding Agent operates iteratively alongside debugging and validation agents, fostering a collaborative loop that facilitates continuous feedback. This interaction enables the early identification and resolution of syntactic, logical, and functional errors, significantly improving the overall code quality.


\textbf{\noindent{Optimization:}} To further improve the quality of the generated code, the guiding prompts for the Coding Agent are meticulously refined. These prompts include critical elements of PLC coding, such as defining the roles and responsibilities of various code modules, enforcing action constraints for compliance with system safety and operational regulations, and leveraging insights from the detailed plans provided by the Planning Agent. These help the Coding Agent meet not only the structural requirements but also the operational constraints, improving code reliability and practicability.



\subsubsection{Debugging Agent}

The Debugging Agent is essential for analyzing errors and inconsistencies that emerge during the compilation of the ST code. It interprets feedback from the ST compiler and utilizes RAG tools alongside the prompts designed for patch generation to produce the revised code. The revised code is then relayed to the Validation Agent for syntactic and semantic verification. The Debugging Agent consists of three components as follows.
 
 \begin{itemize}
    \item \textbf{Syntactic Fixing Advice Generation.} With the input from the compilation results of the ST compiler within the validation agent, it generates fixing advice based on the syntactic checking results including the error locations and reasons. The fixing employs a structured patch generation process, utilizing a step-by-step Chain-of-Thought (CoT) methodology that analyzes the origin of error code, its info and description respectively. This thorough analysis leads to accurate identification of the reasons of error and generation of fixing advice, significantly enhancing the accuracy and efficiency of error fixing.

    \item \textbf{Semantic Fixing Advice Generation.} This component is to address semantic errors identified during formal verification.
    Similar to the syntactic process, the semantic fixing process is also based on CoT, which guides the LLM to assess the violated property and the reason of violation, pinpoint potential code segments responsible for the property violations, and generate fixing advice. 
    
    \item \textbf{Fixed Code Generation.} This component accepts the fixing advice from the previous processes, along with the generated code and the currently active plan from the Planning Agent, based on which it repairs the code. Its design mirrors that of the Coding Agent, leveraging RAG to access the same comprehensive database of the reference resources. By integrating this information, the Debugging Agent can refine its output before relaying them to the Validation Agent.
 \end{itemize}

\textbf{Optimization}
The Debugging Agent serves as a critical module for rectifying code errors. Inspired by existing work in automated code repair~\cite{xia2023keep}, which utilizes the generated dialogue history and real-time feedback for corrective suggestions, we develop a comprehensive workflow for patch analysis and generation within the Debugging Agent. 
We adopt the CoT methodology to guide code repair through the patch generation process. Compared to the traditional fine-tuning methods for code repair, our approach enhances the effectiveness and adaptability of the repair process.
By encouraging the LLM to reflect on error messages, relevant code lines, and test names, the repair process becomes more intuitive and responsive. 
This structured reflection fosters a deeper understanding of the issues at hand, resulting in more precise and effective fixes.

Furthermore, providing contextual information, such as formal verification properties and validation outcomes helps the code repair model generate more accurate and targeted corrections. This strategy significantly enhances the repair process, particularly for errors that cannot be resolved solely through an examination of the source code.

The use of RAG for code generation also proves effective during the debugging phase. Employing previously generated patches in a self-RAG context can improve the effectiveness of repair, capitalizing on the wealth of information contained in the generated outputs. By continuously refining and validating code through this advanced debugging framework, the Debugging Agent not only enhances the quality of the final output but also optimizes the overall development workflow.

\subsubsection{Validation Agent}

The Validation Agent is responsible for verifying the functional correctness of ST code.
Different from LLM4PLC~\cite{fakih2024llm4plc}, it offers code-level verification for the ST code, which is necessary for the industrial control systems. 
The agent is composed of several specialized sub-components, each tasked with specific validation responsibilities:

\begin{itemize}

\item \textbf{ST Code Compilation.} The first step in the validation process is to run the received ST code through a compiler to verify its syntactic correctness. This step ensures the generated code conforms to the syntactical rules of ST.
If the compilation is successful, the process advances to the verification phase. Otherwise, the compiler provides detailed error information, which is then fed into the Debugging Agent for code correction. 
By automating this initial verification step, the Validation Agent helps streamline the debugging cycle and reduces manual intervention.

\item \textbf{Property Generation.} In cases where formal specifications are not explicitly provided by the user, the agent uses a subcomponent that leverages an LLM-driven mechanism to generate a set of formal specifications automatically. 
These specifications are derived from user input, industry standards, and general safety requirements, and they are structured to meet the format required by formal verification tools. This process help make the validation process efficient, especially in complex or large-scale systems where manual specification writing can be time-consuming and error-prone.

\item \textbf{Translation-based ST Code Analysis.} 
This subcomponent translates the generated ST code into formats or languages compatible with formal verification tools such as SMV or CBMC~\cite{kroening2014cbmc}. The translation process is either managed by LLM-driven agents or by specialized tools like PLCverif~\cite{darvas2015PLCverif}. 

\end{itemize}

\textbf{\noindent{Optimization}}
One primary challenge with LLM-guided verification is the potential of inconsistencies between the original ST code and the translated version used for formal verification. Also, automatically translated code may lead to state explosion, a common problem in model checking where the state space grows exponentially, making verification computationally expensive or infeasible. To address these, we integrate advanced translation-based verification tools like PLCverif, which has been optimized for translating ST code into SMV or CBMC formats for model checking and bounded model checking (BMC), and may help alleviate the problem of state explosion.
Our design of the Validation Agent allows for easy incorporation of new verification tools as they become available, future-proofing the framework and ensuring that it remains adaptable to advances in the field of formal verification.

%

Another challenge is that in real-world applications, users often struggle to define formal properties for verification. 
To address this, we develop a specification generation tool within the Validation Agent. It helps users automatically generate formal properties in a format suitable for verification tools, leveraging the chain-of-thought (CoT) methodology to guide the process.
This automated approach not only provide convenience for users but also ensures that the generated properties are tailored to formal verification.



\section{Experimental Evaluation}


To systematically evaluate the effectiveness, efficiency, and other key aspects of our framework for PLC code generation, we design a series of experiments aimed at answering the following research questions (RQs):
\begin{itemize}
\item RQ1: Can Agents4PLC generate PLC code more effectively compared to the existing approaches?
\item RQ2: How efficient Agent4PLC is in PLC code generation? 
\item RQ3: How effective are the designs in agents, e.g., RAG and prompt design in the Coding Agent? 
\item RQ4: How useful is the generated code in practical PLC production environments?   
\end{itemize}

\vspace{1mm}
\noindent{\textbf{Benchmark Construction.}} To accurately evaluate our Agents4PLC system, we constructed the first benchmark dataset focused on the task of generating ST code from natural language specification, and we assess the correctness of code samples automatically through formal verification methods. This dataset comprises 23 programming tasks along with corresponding formal verification specifications, including 58 properties in easy set with 53 non-trivial property over 16 easy programming tasks and 43 in medium set with 38 non-trivial property over 7 medium programming tasks, where trivial property means "assertion" property without corresponding assertion sentences in ST code for reference. These programming problems cover various aspects of industrial programming including Logical Control, Mathematical Operations, Real-time Monitoring,
Process Control and other fields, which effectively simulate the genuine requirements found in industrial control systems.

\vspace{1mm}
\noindent{\textbf{Base LLMs and Retrieval Model.}}
For a comprehensive evaluation of Agents4PLC against different base models, we investigated the capabilities of several popular code LLMs. In particular, we adopt CodeLlama 34B\cite{roziere2023code}, DeepSeek V2.5\cite{guo2024deepseek}, GPT-4o\cite{GPT-4o} and GPT-4o-mini\cite{GPT-4o-mini} to evaluate our method. For the retrieval model, we utilize text-embedding-ada-002, an advanced model developed by OpenAI.

\vspace{1mm}
\noindent{\textbf{Evaluation Metrics.}}
Following previous work\cite{fakih2024llm4plc}, we employ the 1) pass rate, 2) syntax compilation success rate and 3) verification success rate (or verifiable rate) as metrics to evaluate the effectiveness of code generation.
The syntax compilation success rate serves as a preliminary validation of the syntactic correctness of the generated ST code. A high syntax compilation success rate indicates
fewer syntax errors, thereby reducing subsequent debugging efforts.
The pass rate is derived from the pass@k metric, where the model is considered successful if at least one of the \textit{k} generated results that not only compile successfully but also adhere to the specified functional requirements of the PLC program, which is the most challenging task.
The verifiable rate quantifies the proportion of the generated ST code segments that can pass syntax check of target verification language, e.g. nuXmv or cbmc. A high verifiable rate reflects the effectiveness of the approach in generating executable code for verification. We calculate it as the proportion of code segments that successfully compile out of the total number of generated code segments.



\section{Results}

\subsection{RQ1: Effectiveness study} 
In this experiment, we systematically compare our framework with other code generation frameworks based on several base LLMs for generating reliable ST code. 
Our evaluation is based on a set of benchmark cases designed to reflect varying levels of complexity, from relatively straightforward control sequences (labeled as ``Easy") to more sophisticated logic processes (labeled as ``Medium"). 
Note that LLM4PLC is designed to be a half-automated framework with human interactions, we write an extra automation program to drive the components of the LLM4PLC framework. More experiment details are included in our github link \footnote{\href{https://github.com/Luoji-zju/Agents4PLC_release}{https://github.com/Luoji-zju/Agents4PLC\_release}}
and our site.\footnote{\href{https://hotbento.github.io/Agent4PLC/}{https://hotbento.github.io/Agent4PLC/}}

The models evaluated in this experiment include CodeLlama 34B, GPT-4o, GPT-4o-mini, and DeepSeek V2.5. Among these, the CodeLlama 34B model is run on a single NVIDIA A800 80GB PCIe GPU with pre-trained LoRAs from the LLM4PLC framework, while the other models are tested via their respective online APIs. Each model is provided with user requirements, code skeletons, and natural language specifications to generate ST code. This setup mirrors real-world coding scenarios where the models function as code generation agents without detailed control over the underlying logic design. Additionally, to assess the potential of non-specialized models for PLC code generation, we also evaluated the performance of the ChatDev framework\cite{chatdev} with GPT-4o base model on our benchmark.

The experiment results is shown in Table \ref{tab:experiment_1}, which illustrates the performance of different frameworks based on different models across both ``Easy" and ``Medium" benchmark levels. The table presents the pass rates for both compilation and verification stages, with the format X Y Z\%, where X represents the number of successful passes on corresponding metrics, Y indicates the total number of programming problems, and Z\% denotes the pass rate percentage. For instance, LLM4PLC/GPT-4o achieves a syntax compilation pass rate of 14 16 87.5\%, meaning the generated result of LLM4PLC on the GPT-4o model successfully compiles for 14 out of 16 programming problems, yielding an 87.5\% success rate. We record syntax compilation success rate, verifiable rate and pass rate for both systems, where ``verifiable rate" means the framework can generate verifiable model for at least 80\% of the given properties, and ``pass rate" means that least 80\% of the generated code can compile.

The results highlight the effectiveness of our Agents4PLC framework which outperforms other software development frameworks across different benchmark levels. In the ``Easy'' category, the performance of LLM4PLC on different models demonstrates notable variability. The GPT-4o-mini model achieves the highest syntax compilation pass rate of 93.8\%, successfully compiling 15 out of 16 code segments. In contrast, the CodeLlama 34B model, based on the original experimental setting has the lowest performance, with a syntax compilation pass rate of only 68.8\%. Notably, the verifiable rate and pass rate for all LLM4PLC models are significantly low, with both GPT-4o and DeepSeek V2.5 recording a verifiable rate of 12.5\% on the Easy problems. Only DeepSeek V2.5 achieves a pass rate of 12.5\% on these problems, indicating that the LLM-based automative generation of SMV models requires further improvement. 

On the contrary, Agents4PLC framework exhibits more consistent performance across different models in the Easy category. 
Except for CodeLlama 34B, they all achieve a syntax compilation pass rate of 100\%, successfully compiling all generated code for the programming problems. 
\textbf{Our Agents4PLC framework across different models achieves a maximum verifiable rate of 68\% and a pass rate of 50\% with the GPT-4o model, indicating the superior capability of our framework in generating verifiable code.}

For the Medium benchmark level, both systems maintain their respective performance patterns, with \emph{Agents4PLC consistently outperforming LLM4PLC}. All models from Agents4PLC manage to achieve a syntax compilation rate of 100\% on ``Medium'' problems, contrasting sharply with the variable results from LLM4PLC, where the highest syntax compilation pass rate is again recorded by GPT-4o and GPT-4o-mini at 57.1\%. \textbf{Our Agents4PLC framework achieves a maximum verifiable rate of 42.3\% with GPT-4o and DeepSeek V2.5, and a maximum pass rate of 28.6\% with GPT-4o, demonstrating that our framework can effectively handle tasks involving complex coding problems.}

We also conduct experiments on ST code generation using ChatDev, a general-purpose software development platform based on multi-agent systems. This software development framework achieve 43.8\% syntax compilation, 43.8\% verifiable rate, 43.8\% pass rate on the Easy category and 28.6\% syntax compilation, 14.3\% verifiable rate, and 28.6\% pass rate on medium category, showing that most compilable code from ChatDev are correct in semantics. However, it is also worth noting that despite explicitly prompting the ChatDev framework to generate ST code, \textit{it occasionally produces code in unrelated languages, such as Python or C++}. This highlights a limitation of general-purpose code generation frameworks in specialized industrial domains.

\begin{table*}[htbp]
\caption{Efficiency evaluation: ST code generation times over all attempts passing syntactic compilation}
\centering
\begin{tabular}{|m{0.09\linewidth}|m{0.11\linewidth}|m{0.108\linewidth}|m{0.108\linewidth}|m{0.108\linewidth}|m{0.108\linewidth}|m{0.108\linewidth}|m{0.108\linewidth}|}
\hline
\multicolumn{2}{|c|}{\textbf{Benchmark Level}}  & \multicolumn{3}{|c|}{\textbf{Easy Problems}} & \multicolumn{3}{|c|}{\textbf{Medium Problems}} \\
\hline
 \textbf{Framework} &  \textbf{Base Model} & \makecell[c]{1 time} & \makecell[c]{2 time}  & \makecell[c]{3 or more} & \makecell[c]{1 time} & \makecell[c]{2 time}  & \makecell[c]{3 or more} 
 \\
\hline

\cline{2-8}
\multirow{4}{*}{\textbf{LLM4PLC}} & \textbf{CodeLlama 34B} & 8 \ 11 \ 72.7\%   & 0 \ 11 \ 0.0\%   & 3 \ 11 \ 27.3\%   & 2 \ 4 \ 50.0\%   & 2 \ 4 \ 50.0\%   & 0 \ 4 \ 0.0\%  \\
\cline{2-8}
& \textbf{DeepSeek V2.5} & 12 \ 13 \ 92.3\%  & 1 \ 13 \ 7.7\%   & 0 \ 13 \ 0.0\%   & 6 \ 7 \ 85.7\%   & 1 \ 7 \ 14.3\%   & 0 \ 7 \ 0.0\%  \\
\cline{2-8}
 & \textbf{GPT-4o}        & 13 \ 14 \ 92.9\%  & 1 \ 14 \ 7.1\%   & 0 \ 14 \ 0.0\%   & 3 \ 4 \ 75.0\%   & 1 \ 4 \ 25.0\%   & 0 \ 4 \ 0.0\%  \\
\cline{2-8}
 & \textbf{GPT-4o-mini}   & 15 \ 15 \ 100.0\% & 0 \ 15 \ 0.0\%   & 0 \ 15 \ 0.0\%   & 3 \ 4 \ 75.0\%   & 0 \ 4 \ 0.0\%    & 1 \ 4 \ 25.0\% \\
\hline
\cline{2-8}
\multirow{4}{*}{\textbf{Agents4PLC}} & \textbf{CodeLlama 34B} & 4 \ 5 \ 80.0\%   & 1 \ 5 \ 20.0\%  & 0 \ 5 \ 0.0\%    & 1 \ 1 \ 100.0\%  & 0 \ 1 \ 0.0\%    & 0 \ 1 \ 0.0\%  \\
\cline{2-8}
& \textbf{DeepSeek V2.5} & 16 \ 16 \ 100.0\%& 0 \ 16 \ 0.0\%  & 0 \ 16 \ 0.0\%   & 7 \ 7 \ 100.0\%  & 0 \ 7 \ 0.0\%    & 0 \ 7 \ 0.0\%  \\
\cline{2-8}
& \textbf{GPT-4o}        & 16 \ 16 \ 100.0\%& 0 \ 16 \ 0.0\%  & 0 \ 16 \ 0.0\%   & 7 \ 7 \ 100.0\%  & 0 \ 7 \ 0.0\%    & 0 \ 7 \ 0.0\%  \\
\cline{2-8}
 & \textbf{GPT-4o-mini}   & 16 \ 16 \ 100.0\%& 0 \ 16 \ 0.0\%  & 0 \ 16 \ 0.0\%   & 7 \ 7 \ 100.0\%  & 0 \ 7 \ 0.0\%    & 0 \ 7 \ 0.0\%  \\
\hline

\end{tabular}
\\[6pt] 

\footnotesize{Note: The format of data represents the number for corresponding generation times / total cases passing syntactic compilation / ratio.}
\label{tab:experiment_2}
\end{table*}

\begin{table*}[htbp]
\caption{Ablation Experiment with Designs on Coding and Fixing Agents}
\centering
\begin{tabular}{|m{0.09\linewidth}|m{0.21\linewidth}|m{0.1\linewidth}|m{0.095\linewidth}|m{0.095\linewidth}|m{0.085\linewidth}|m{0.085\linewidth}|m{0.085\linewidth}|}
\hline
\multicolumn{2}{|c|}{\textbf{Benchmark Level}}  & \multicolumn{3}{|c|}{\textbf{Easy Problems}} & \multicolumn{3}{|c|}{\textbf{Medium Problems}} \\
\hline
\textbf{Framework} & \textbf{Base Model} & \makecell[c]{Syntax \\ Compilation} & \makecell[c]{Pass Rate}  & \makecell[c]{Verifiable  Rate} & \makecell[c]{Syntax \\ Compilation} & \makecell[c]{Pass  Rate}  & \makecell[c]{Verifiable  Rate} \\
\hline
\multirow{5}{*}{\textbf{Coding Agent}} & \textbf{One-shot + RAG + Syntax Hint} & 16 \ 16 \ 100.0\% & 8 \ 16 \ 50.0\% & 11 \ 16 \ 68.8\% & 7 \ 7 \ 100.0\% & 2 \ 7 \ 28.6\% & 3 \ 7 \ 42.9\% \\ 
\cline{2-8}
& \textbf{One-shot + Syntax Hint} & 16 \ 16 \ 100.0\% & 11 \ 16 \ 68.8\% & 12 \ 16 \ 75.0\% & 7 \ 7 \ 100.0\% & 1 \ 7 \ 14.3\% & 3 \ 7 \ 42.9\% \\ 
\cline{2-8}
& \textbf{One-shot + RAG} & 16 \ 16 \ 100.0\% & 6 \ 16 \ 37.5\% & 9 \ 16 \ 56.2\% & 7 \ 7 \ 100.0\% & 1 \ 7 \ 14.3\% & 1 \ 7 \ 14.3\% \\ 
\cline{2-8}
& \textbf{One-shot} & 16 \ 16 \ 100.0\% & 6 \ 16 \ 37.5\% & 9 \ 16 \ 56.2\% & 7 \ 7 \ 100.0\% & 0 \ 7 \ 0.0\% & 1 \ 7 \ 14.3\% \\ 
\cline{2-8}
& \textbf{Zero-shot} & 16 \ 16 \ 100.0\% & 7 \ 16 \ 43.8\% & 8 \ 16 \ 50.0\% & 7 \ 7 \ 100.0\% & 0 \ 7 \ 0.0\% & 1 \ 7 \ 14.3\% \\ 
\hline
\textbf{Debugging Agent} & \textbf{Without CoT / Patch Template} & 16 \ 16 \ 100.0\% & 10 \ 16 \ 62.5\% & 10 \ 16 \ 50.0\% & 7 \ 7 \ 100.0\% & 0 \ 7 \ 0.0\% & 1 \ 7 \ 14.3\% \\ 
\hline
\end{tabular}
\\[6pt] 

\footnotesize{Note: The format of data represents number of successful passes on corresponding
metrics / total number of programming problems / passing rate.}
\label{tab:ablation_experiment}
\end{table*}


\subsection{RQ2: Efficiency study}
\label{sec:V.B}

To evaluate the efficiency of our framework in generating verifiable st code, we measured the time taken for each model to successfully generate a syntactically correct and semantically verified ST code. Considering in our validation process, PLCverif is not supported for efficiency evaluation, we categorize the results based on how many attempts it takes for each model to pass the syntactic compilation stage. The categories are as follows:

\begin{itemize}
\item \textbf{1 Attempt:} The code passes syntactic compilation stage on the first try.
\item \textbf{2 Attempts:} The code passes syntactic compilation stage on the second attempt after a failure.
\item \textbf{3 or more than 3 Attempts:} The framework takes more than three attempts to succeed.
\end{itemize}

The experimental results are presented in Table \ref{tab:experiment_2}, where the data format represents the number of generation attempts required / total cases passing syntactic compilation / ratio. The experimental setup is identical to that of Experiment 1.

The results demonstrate that our framework, when paired with the base models DeepSeek V2.5, GPT-4o, and GPT-4o-mini, consistently achieved successful ST code generation in a single attempt, regardless of whether the problems are classified as easy or medium. In contrast, the CodeLlama 34B model within our framework exhibited instances of requiring code repair after the initial attempt. In comparison, the LLM4PLC framework showed multiple instances across all tested models where two or more attempts are necessary to produce compilable code. This stark contrast in performance underscores that \textbf{our Agents4PLC framework not only delivers higher code generation success rates but also significantly improves efficiency, particularly when compared to LLM4PLC, which required more frequent code corrections.}



\subsection{RQ3: Ablation study}

The ablation study aims to evaluate the influence of specific design choices within our framework on the performance of code generation.
Our ablation experiments are organized around two primary domains:
\begin{itemize}

    \item  \textbf{Coding Agent}: We investigate the effects of three significant enhancements: syntax hints in prompts, retrieval-augmented generation (RAG), and one-shot prompting on the ST code generation process. The following configurations are examined, with each enhancement systematically removed to assess its impact on the pass rates: 

    \begin{itemize}
        \item  \textbf{Full Configuration}: The complete framework incorporating syntax hints, RAG, and one-shot prompting.
        \item  \textbf{Intermediate Configuration without RAG}: The framework utilizing one-shot prompting and syntax hints, but excluding RAG.
        \item  \textbf{Intermediate Configuration without Syntax hint}: The framework utilizing simple one-shot prompting and RAG, but excluding syntax hints.
        \item  \textbf{Simplified Configuration}: A streamlined version using only plain one-shot prompting.
        \item  \textbf{Baseline Configuration}: The foundational setup with zero-shot prompting and no supplementary aids (syntax hints and RAG).
    \end{itemize}

     \item  \textbf{Debugging Agent}: In this segment, we assess the importance of two critical components of the Debugging agent: chain-of-thought (CoT) reasoning and patch templates. We compare the framework's performance when both components are disabled, effectively operating without CoT reasoning and patch templates. This allows us to quantify the degradation in code correction capabilities resulting from their absence.


Table \ref{tab:ablation_experiment} summarizes the findings from this ablation study, showcasing the ST code verification pass rates and compilation pass rates across various configurations. The configuration of experiment and table content is similar to those in effectiveness study, including metrics and data format. 
The configuration that on Easy Problem set, the setting combining One-shot and Syntax Hint yields the highest overall pass rates (68.8\%) and verifiable rates (75.0\%). However, the results for medium problems mirror the full configuration, One-shot + RAG + Syntax Hint configuration, demonstrate superior performance with 28.6\% pass rate and 42.9\% verifiable rate). The impact for each configuration analysis is:

\begin{itemize}
    \item  \textbf{RAG}: The application of RAG provides noticeable improvements for medium-level problems, but its impact on easy problems is less pronounced, and in some cases, it may even have unintended negative effects.
    \item  \textbf{Syntax Hint}: Detailed syntax hints significantly enhance the effectiveness of code generation, especially for easy problems. The provision of syntax guidance helps improve both the syntactical correctness and the overall quality of generated code.
    \item  \textbf{One shot}: In our experiments, we observe that one-shot prompting do not lead to a substantial improvement in performance. This is likely because the one-shot method merely provides a reference ST code template, which has limited effectiveness in improving the overall quality of code generation.
    \item  \textbf{CoT in Debugging agent}: For easy problems, the results show no significant difference between using the standard debugging agent and that with CoT. However, for medium-level problems, the removal of CoT noticeably reduce the framework's performance, indicating that CoT is crucial for handling more complex debugging tasks.
\end{itemize}

\end{itemize}

The results indicate that the inclusion of syntax hints improves performance across all metrics, while the impact of one-shot prompting remains unclear. However, the effects of more advanced optimization techniques, such as RAG and CoT, warrant further investigation. These methods significantly enhance the framework's ability to handle complex problems, but when not carefully designed, they may interfere with the reasoning process on simpler tasks. Additionally, considering Section \ref{sec:V.B}, our framework can often generate correct code in a single attempt. As a result, the current experiments may not fully capture the effectiveness of the debugging agent.

\subsection{RQ4: Case study in practical control environment}

To evaluate how Agent4PLC performs in practical industrial control environment, we conduct case studies utilizing the UWinTech Control Engineering Application Software Platform, developed by Hangzhou UWNTEK Automation System Co., Ltd. UWinTech is a software package designed for using the UW series control system~\cite{Uwin}. It integrates a wide range of functionalities, including on-site data collection, algorithm execution, real-time and historical data processing, alarm and safety mechanisms, process control, animation display, trend curve analysis, report generation, and network monitoring. The engineer station configuration software, operator station real-time monitoring software, and on-site control station real-time control software operate on different levels of hardware platforms. Its components interact via control networks and system networks, coordinating the exchange of data, management, and control information to ensure the successful execution of various functions within the control system.

Our experiments involve several steps: first, we constructed the operation station and the control station, then we configured the attributes of the monitoring and control points.
These points are linked to the simulation model, and the generated ST code from Agent4PLC were uploaded to the control station to conduct the experiment. 
We perform four control tasks: site monitoring and alarm light flashing, low voltage limit and motor start/stop, temperature and pressure monitoring, and specific node delay monitoring.



\begin{figure}
  \vspace{-3mm}
  \centering
  \includegraphics[width=.98\linewidth]{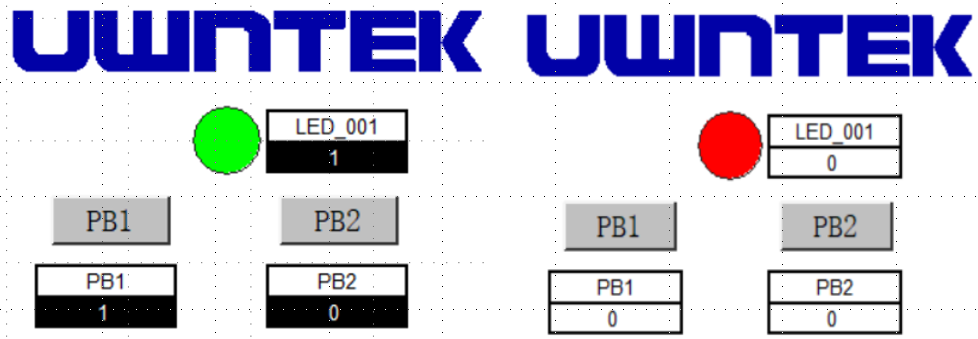}
  \caption{LED control with Agents4PLC.}
  \label{fig:case1}
\end{figure}

Figure~\ref{fig:case1} denotes an LED Control task.
The prompt is: design a PLC function block named LED-Control using structured text (ST) code. The function block should contain two Boolean input variables (PB1 and PB2) and one Boolean output variable (LED). Implement logical operations such that LED is assigned the value resulting from a logical AND operation between PB1 and the negation of PB2. The function block should represent a simple control system where PB1 and PB2 determine whether LED is on or off based on the specified logic.
The results show that with the generated ST code uploaded, when PB1 is true ($1$) and PB2 is false ($0$), namely the the AND operation between PB1 and the negation of PB2 output $1$, the light is green, and otherwise, it is red. 
This complies with the requirement in the prompt.

\begin{figure}
  \centering
  \includegraphics[width=.95\linewidth]{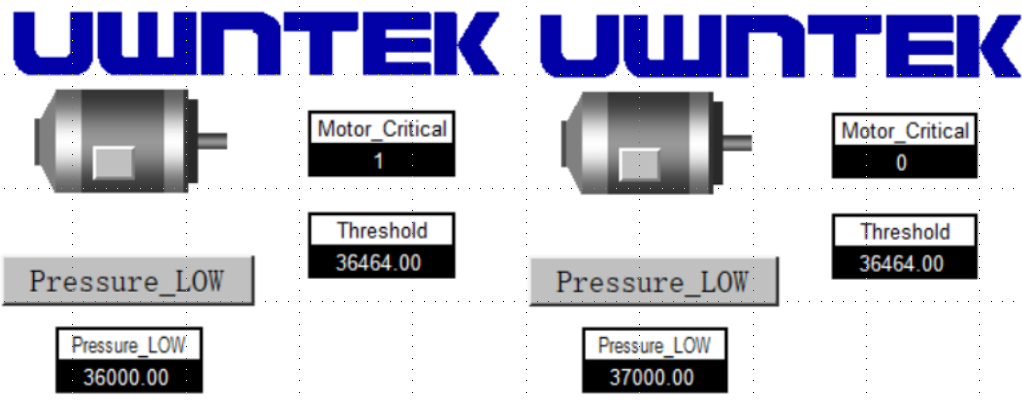} 
  \caption{Motor triggering control with Agents4PLC.}
  \vspace{-2mm}
  \label{fig:case2}
\end{figure}

Figure~\ref{fig:case2} denotes a motor control task.
The Prompt is: Design a PLC function block in Structured Text (ST) that evaluates whether the critical motor should be triggered based on the given low pressure value compared to a threshold of $36464$. The state of Motor-Critical is determined based on this evaluation.
The results show that when the input voltage is below the threshold 36464, the motor stops; otherwise, the motor starts.

\begin{figure}
  \centering
  \includegraphics[width=\linewidth]{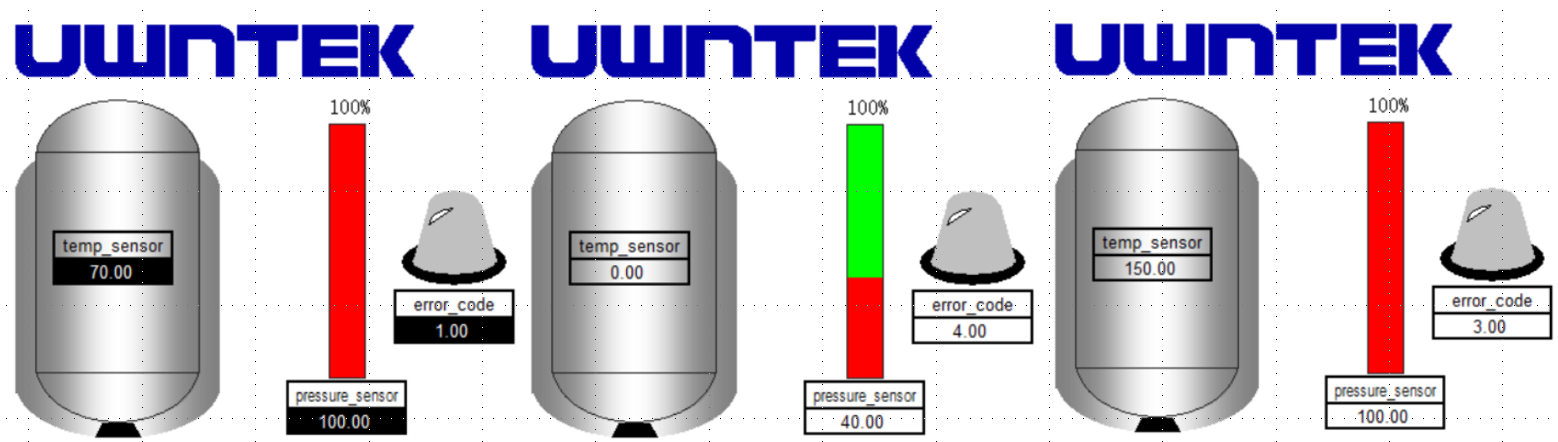} 
  \caption{Temperature update with Agents4PLC.}
  \vspace{-5mm}
  \label{fig:case3}
\end{figure}

Figure~\ref{fig:case3} denotes a temperature and relay update task.
The prompt is: Design a PLC program using structured text (ST code) that incorporates pressure sensors, temperature sensors, relays, counters, error codes, and error flags as inputs. The program must loop through the pressure sensor, adjust the temperature sensor based on specific conditions, and update the relay status according to the value of GT1-OUT. Ensure that the program checks for conditions to avoid overflow and maintains the error flag state. Return a Boolean value indicating the completion of the operation.
The results show that low or high limit alarms and different error codes are provided when pressure and temperature are abnormal. When an abnormality occurs, the alarm light will flash. 
If the pressure exceeds the upper limit, the error code is 1. 
If the temperature is above the high limit, the error code is 3. If the temperature is below the lower limit, the error code is 4.
These comply with the requirements in the prompt.


 
\begin{table*}[htbp]
\caption{Compilation and Verification Metrics for Different Models}
\centering
\begin{tabular}{|m{0.09\linewidth}|m{0.11\linewidth}|m{0.108\linewidth}|m{0.108\linewidth}|m{0.108\linewidth}|m{0.108\linewidth}|m{0.108\linewidth}|m{0.108\linewidth}|}
\hline
\multicolumn{2}{|c|}{\textbf{Benchmark Level}}  & \multicolumn{3}{|c|}{\textbf{Easy Problems}} & \multicolumn{3}{|c|}{\textbf{Medium Problems}} \\
\hline
 \textbf{Framework} &  \textbf{Base Model} & \makecell[c]{Syntax \\ Compilation} & \makecell[c]{Verifiable  Rate}  & \makecell[c]{Pass Rate} & \makecell[c]{Syntax \\ Compilation} & \makecell[c]{Verifiable  Rate}  & \makecell[c]{Pass  Rate} 
 \\
\hline

\cline{2-8}
\multirow{4}{*}{\textbf{LLM4PLC}}& \textbf{CodeLlama 34B}    & 11 \ 16 \ \ 68.8\%      & 0 \ 16 \ \ 0.0\%     & 0 \ 16 \ \ 0.0\%         & 4 \ 7 \ 57.1\%               & 0 \ 7 \ 0.0\%            & 0 \ 7 \ 0.0\%               \\
\cline{2-8}
& \textbf{DeepSeek V2.5}    & 13 \ 16 \ 81.3\%        & 2 \ 16 \ 12.5\%         & 2 \ 16 \ 12.5\%           & \textbf{7 \ 7 \ 100.0\%}              & 0 \ 7 \ 0.0\%            & 0 \ 7 \ 0.0\%               \\
\cline{2-8}
& \textbf{GPT-4o}           & 14 \ 16 \ 87.5\%        & 0 \ 16 \ 0.0\%          & 2 \ 16 \ 12.5\%           & 4 \ 7 \ 57.1\%               & 0 \ 7 \ 0.0\%            & 0 \ 7 \ 0.0\%               \\
\cline{2-8}
& \textbf{GPT-4o-mini}      & 15 \ 16 \ 93.8\%        & 0 \ 16 \ 0.0\%          & 0 \ 16 \ 0.0\%            & 4 \ 7 \ 57.1\%               & 0 \ 7 \ 0.0\%            & 0 \ 7 \ 0.0\%               \\
\hline
\cline{2-8}
\multirow{4}{*}{\textbf{Agents4PLC}}& \textbf{CodeLlama 34B}    &\ 5 \ 16 \ \  31.3\%               & 2 \ 16 \ 12.5\%      & 1 \ 16 \ 6.3\%         & 1 \ 7 \ 14.3\%               & 0 \ 7 \ 0.0\%            & 0 \ 7 \ 0.0\%               \\
\cline{2-8}
& \textbf{DeepSeek V2.5}    & \textbf{16 \ 16 \ 100.0\%}        &\textbf{10 \ 16 \ 62.5\%}    & \textbf{7 \ 16 \ 43.8\%}       & \textbf{7 \ 7 \ 100.0\%}                  & \textbf{3 \ 7 \ 42.9\%}      & \textbf{1 \ 7 \ 14.3\%}              \\
\cline{2-8}
& \textbf{GPT-4o}           & \textbf{16 \ 16 \ 100.0\%}              & \textbf{11 \ 16 \ 68.8\%}   & \textbf{8 \ 16 \ 50.0\%}         & \textbf{7 \ 7 \ 100.0\%}           & \textbf{3 \ 7 \ 42.9\%}   & \textbf{2 \ 7 \ 28.6\%}         \\
\cline{2-8}
& \textbf{GPT-4o-mini}      & \textbf{16 \ 16 \ 100.0\%}              & \textbf{8 \ 16 \ 50.0\%}     & \textbf{7 \ 16 \ 43.8\%}        & \textbf{7 \ 7 \ 100.0\%}       & 1 \ 7 \ 14.3\%     & 0 \ 7 \ 0.0\%       \\
\hline
\textbf{ChatDev} & \textbf{GPT-4o}      & 7 \ 16 \ 43.8\%       & NA       & NA            &  2 \ 7 \ 28.6\%               & NA           & NA               \\
\hline
\end{tabular}
\\[6pt] 

\footnotesize{Note: The format of data represents number of successful passes on corresponding
metrics / total number of programming problems / passing rate.}
\label{tab:experiment_1}
\end{table*}
\section{Related works}
There are significantly increasing interests from both the industry and academia in LLM-based code generation, including those LLMs specifically designed for code generation such as DeepSeek-Coder\cite{guo2024deepseek}, StarCoder \cite{li2023starcoder}, and CodeLlama \cite{roziere2023code}. 
However, most of them focus on high-level languages such as C, Python and only a very small portion considers PLC code in control engineering.
Here, we mainly review those works considering PLC code from the aspects of its generation, testing and verification.
We also seperatly review the existing LLM-based multi-agent frameworks for code generation.
\subsection{LLM-based Automated PLC Code Generation}
There exist a few works studying the LLM-based generation of PLC programming code like ST. 
For instance, Koziolek et al.~\cite{koziolek2023chatgpt} create 100 prompts across 10 categories to evaluate the ability of the existing LLMs to produce syntactically correct control logic code in ST language.
Later, they introduce a retrieval-augmented generation method~\cite{koziolek2024llm-retrieval} and an image-recognition-based generation method~\cite{koziolek2024llm-image}. These methods integrate proprietary function blocks into the generated code and utilize GPT-4 Vision to generate control logic code for industrial automation from Piping-and-Instrumentation Diagrams (P\&IDs), respectively.
However, they do not consider testing or verification of the generated code, and thus cannot ensure code correctness.
Fakih et al. \cite{fakih2024llm4plc} introduce an LLM-based PLC code generation pipeline named LLM4PLC, which integrates the fine-tuned LLMs with external verification tools.
Though it takes formal verification into consideration, it only achieves design-level verification (instead of the ST code level) and the generation process can not achieve full automation.
Witnessing these limitations of existing works, we aim to achieve closed-loop and fully automated PLC code generation and verification with our designed multi-agent system Agents4PLC, paving a way for evolving, efficient, trustworthy and intelligent coding for industrial control systems. 

Our work is also inspired by the recent trend in code generation approaches which rely on the cooperation of LLM-based agents.
ChatDev \cite{hou2023large}, a virtual software development company composed of multiple agents, features clearly defined roles and divisions of labor, aiming to collaboratively generate high-quality software code. However, ChatDev still shows limitations when it comes to generating software code for industrial control systems without formal verification support.
Mapcoder~\cite{islam2024mapcoder} consists of four LLM-based agents for the tasks of recalling relevant examples, planning, code generation, and debugging respectively, relying on multi-agent prompting for code generation.
AutoSafeCoder~\cite{nunez2024autosafecoder} consists of three agents responsible for code generation, static analysis and fuzzing to detect runtime errors respectively.
AgentCoder~\cite{huang2023agentcoder} consists of three agents responsible for code generation and refinement, test case generation, test execution and feedback reporting respectively. 
These approaches all focus on Python language which is fundamentally different to ST for PLC. Agents4PLC is the first LLM-agent-based system covering the whole lifecycle of ST control engineering.


\subsection{PLC Code Testing and Verification} 

Due to the fact that PLC programming is often performed in low-level programming languages, which typically use bitwise and boolean operations, it becomes very difficult to understand and debug PLC programs. This increases the need for testing and verification of PLC programs~\cite{ovatman2016overview, singh2023taxonomy}. Existing methods for the automated generation of PLC test cases mainly include symbolic execution~\cite{shi2024automated}, concolic testing \cite{bohlender2016concolic} and search-based techniques~\cite{ebrahimi2023pylc}. However, these approaches can produce test cases that are difficult to maintain, making them challenging to use. 
Koziolek et al.~\cite{koziolek2024automated} propose to automatically generate PLC test cases as a csv file by querying an LLM with a prompt to synthesize code test cases. 
And they found in experiments that many generated test cases contain incorrect assertions and require manual correction.
The multi-agent based generation approaches Mapcoder~\cite{islam2024mapcoder}, AutoSafeCoder\cite{nunez2024autosafecoder} and AgentCoder~\cite{huang2023agentcoder} all consider code testing, but ignore formal verification of the generated code.

In terms of PLC code verification, there exist tools like nuXmv \cite{cavada2014nuxmv} and PLCverif \cite{darvas2015PLCverif} applicable for functional verification of ST code, which are serving as the backend verifier in Agents4PLC. 
Besides, several recent works aim to establish formal semantics for IEC 61131-3 languages like ST with more recent language framework like K framework \cite{wang2023k,lee2024formal}, facilitating testing or verification of ST code \cite{lee2022bounded}.
\section{Conclusion and Future Works}

In this paper, we presented Agents4PLC, the first LLM-based multi-agent framework that addresses the critical challenges of automated Programmable Logic Controller (PLC) code generation and verification. By establishing a comprehensive benchmark that transitions from natural language requirements to formal specifications, we laid the groundwork for future research in the field of PLC code generation. Our framework not only emphasizes code-level verification and full automation, 
but also is flexible to incorporate various base code generation models and development modules.
Extensive evaluation demonstrates that Agents4PLC significantly outperforms the previous approaches, achieving high automation and verifiability for PLC code generation.

In the future, we plan to expand the framework to support additional PLC programming languages and standards to enhance its applicability across various industrial contexts. Moreover, we will explore user feedback mechanisms within the multi-agent system to help refine the generated code and the agents based on real-world usage, thereby further enhancing its usability and reliability.

\bibliography{Agents4ICS}

\begin{thebibliography}{10}
\providecommand{\url}[1]{#1}
\csname url@samestyle\endcsname
\providecommand{\newblock}{\relax}
\providecommand{\bibinfo}[2]{#2}
\providecommand{\BIBentrySTDinterwordspacing}{\spaceskip=0pt\relax}
\providecommand{\BIBentryALTinterwordstretchfactor}{4}
\providecommand{\BIBentryALTinterwordspacing}{\spaceskip=\fontdimen2\font plus
\BIBentryALTinterwordstretchfactor\fontdimen3\font minus \fontdimen4\font\relax}
\providecommand{\BIBforeignlanguage}[2]{{%
\expandafter\ifx\csname l@#1\endcsname\relax
\typeout{** WARNING: IEEEtranS.bst: No hyphenation pattern has been}%
\typeout{** loaded for the language `#1'. Using the pattern for}%
\typeout{** the default language instead.}%
\else
\language=\csname l@#1\endcsname
\fi
#2}}
\providecommand{\BIBdecl}{\relax}
\BIBdecl

\bibitem{Uwin}
``Uwintech control engineering application software platform,'' \url{https://www.uwntek.com/product/2.html}, accessed: 2024-10-10.

\bibitem{MATIEC}
\BIBentryALTinterwordspacing
``Matiec,'' 2017. [Online]. Available: \url{https://github.com/nucleron/matiec}
\BIBentrySTDinterwordspacing

\bibitem{GPT-4o}
\BIBentryALTinterwordspacing
``Gpt-4o,'' 2024. [Online]. Available: \url{https://platform.openai.com/docs/models/gpt-4o}
\BIBentrySTDinterwordspacing

\bibitem{GPT-4o-mini}
\BIBentryALTinterwordspacing
``Gpt-4o-mini,'' 2024. [Online]. Available: \url{https://platform.openai.com/docs/models/gpt-4o-mini}
\BIBentrySTDinterwordspacing

\bibitem{LangGraph}
\BIBentryALTinterwordspacing
``Langgraph,'' 2024. [Online]. Available: \url{https://github.com/langchain-ai/langgraph}
\BIBentrySTDinterwordspacing

\bibitem{RuSTy}
\BIBentryALTinterwordspacing
``Rusty,'' 2024. [Online]. Available: \url{https://github.com/PLC-lang/rusty}
\BIBentrySTDinterwordspacing

\bibitem{bohlender2016concolic}
D.~Bohlender, H.~Simon, N.~Friedrich, S.~Kowalewski, and S.~Hauck-Stattelmann, ``Concolic test generation for plc programs using coverage metrics,'' in \emph{2016 13th International Workshop on Discrete Event Systems (WODES)}.\hskip 1em plus 0.5em minus 0.4em\relax IEEE, 2016, pp. 432--437.

\bibitem{cavada2014nuxmv}
R.~Cavada, A.~Cimatti, M.~Dorigatti, A.~Griggio, A.~Mariotti, A.~Micheli, S.~Mover, M.~Roveri, and S.~Tonetta, ``The nuxmv symbolic model checker,'' in \emph{Computer Aided Verification: 26th International Conference, CAV 2014, Held as Part of the Vienna Summer of Logic, VSL 2014, Vienna, Austria, July 18-22, 2014. Proceedings 26}.\hskip 1em plus 0.5em minus 0.4em\relax Springer, 2014, pp. 334--342.

\bibitem{chen2021evaluating}
M.~Chen, J.~Tworek, H.~Jun, Q.~Yuan, H.~P. D.~O. Pinto, J.~Kaplan, H.~Edwards, Y.~Burda, N.~Joseph, G.~Brockman \emph{et~al.}, ``Evaluating large language models trained on code,'' \emph{arXiv preprint arXiv:2107.03374}, 2021.

\bibitem{chen2024supersonic}
Z.~Chen, S.~Fang, and M.~Monperrus, ``Supersonic: Learning to generate source code optimizations in c/c++,'' \emph{IEEE Transactions on Software Engineering}, 2024.

\bibitem{corso2024generating}
V.~Corso, L.~Mariani, D.~Micucci, and O.~Riganelli, ``Generating java methods: An empirical assessment of four ai-based code assistants,'' in \emph{Proceedings of the 32nd IEEE/ACM International Conference on Program Comprehension}, 2024, pp. 13--23.

\bibitem{darvas2015PLCverif}
D.~Darvas, E.~Blanco~Vinuela, and B.~Fern{\'a}ndez~Adiego, ``Plcverif: A tool to verify plc programs based on model checking techniques,'' in \emph{Proceedings of the 15th International Conference on Accelerator and Large Experimental Physics Control Systems}.\hskip 1em plus 0.5em minus 0.4em\relax IEEE, 2015, pp. 1--6.

\bibitem{ebrahimi2023pylc}
M.~Ebrahimi~Salari, E.~P. Enoiu, W.~Afzal, and C.~Seceleanu, ``Pylc: A framework for transforming and validating plc software using python and pynguin test generator,'' in \emph{Proceedings of the 38th ACM/SIGAPP Symposium on Applied Computing}, 2023, pp. 1476--1485.

\bibitem{fakih2024llm4plc}
M.~Fakih, R.~Dharmaji, Y.~Moghaddas, G.~Quiros, O.~Ogundare, and M.~A. Al~Faruque, ``Llm4plc: Harnessing large language models for verifiable programming of plcs in industrial control systems,'' in \emph{Proceedings of the 46th International Conference on Software Engineering: Software Engineering in Practice}, 2024, pp. 192--203.

\bibitem{first2022diversity}
E.~First and Y.~Brun, ``Diversity-driven automated formal verification,'' in \emph{Proceedings of the 44th International Conference on Software Engineering}, 2022, pp. 749--761.

\bibitem{guo2024deepseek}
D.~Guo, Q.~Zhu, D.~Yang, Z.~Xie, K.~Dong, W.~Zhang, G.~Chen, X.~Bi, Y.~Wu, Y.~Li \emph{et~al.}, ``Deepseek-coder: When the large language model meets programming--the rise of code intelligence,'' \emph{arXiv preprint arXiv:2401.14196}, 2024.

\bibitem{hong2023metagpt}
S.~Hong, X.~Zheng, J.~Chen, Y.~Cheng, J.~Wang, C.~Zhang, Z.~Wang, S.~K.~S. Yau, Z.~Lin, L.~Zhou \emph{et~al.}, ``Metagpt: Meta programming for multi-agent collaborative framework,'' \emph{arXiv preprint arXiv:2308.00352}, 2023.

\bibitem{hou2023large}
X.~Hou, Y.~Zhao, Y.~Liu, Z.~Yang, K.~Wang, L.~Li, X.~Luo, D.~Lo, J.~Grundy, and H.~Wang, ``Large language models for software engineering: A systematic literature review,'' \emph{ACM Transactions on Software Engineering and Methodology}, 2023.

\bibitem{huang2023agentcoder}
D.~Huang, Q.~Bu, J.~M. Zhang, M.~Luck, and H.~Cui, ``Agentcoder: Multi-agent-based code generation with iterative testing and optimisation,'' \emph{arXiv preprint arXiv:2312.13010}, 2023.

\bibitem{IEC61131-3}
{International Electrotechnical Commission (IEC)}, ``{IEC 61131-3:2013 Programmable controllers - Part 3: Programming languages},'' \url{https://webstore.iec.ch/en/publication/4552}, 2013, edition 3.0, ISBN: 9782832206614.

\bibitem{isbell2001social}
C.~Isbell, C.~R. Shelton, M.~Kearns, S.~Singh, and P.~Stone, ``A social reinforcement learning agent,'' in \emph{Proceedings of the fifth international conference on Autonomous agents}, 2001, pp. 377--384.

\bibitem{islam2024mapcoder}
M.~A. Islam, M.~E. Ali, and M.~R. Parvez, ``Mapcoder: Multi-agent code generation for competitive problem solving,'' \emph{arXiv preprint arXiv:2405.11403}, 2024.

\bibitem{kaelbling1996reinforcement}
L.~P. Kaelbling, M.~L. Littman, and A.~W. Moore, ``Reinforcement learning: A survey,'' \emph{Journal of artificial intelligence research}, vol.~4, pp. 237--285, 1996.

\bibitem{kornaszewski2020use}
M.~Kornaszewski, ``The use of programmable logic controllers in railway signaling systems,'' in \emph{ICTE in Transportation and Logistics 2019}.\hskip 1em plus 0.5em minus 0.4em\relax Springer, 2020, pp. 104--111.

\bibitem{koziolek2024automated}
H.~Koziolek, V.~Ashiwal, S.~Bandyopadhyay \emph{et~al.}, ``Automated control logic test case generation using large language models,'' \emph{arXiv preprint arXiv:2405.01874}, 2024.

\bibitem{koziolek2023chatgpt}
H.~Koziolek, S.~Gruener, and V.~Ashiwal, ``Chatgpt for plc/dcs control logic generation,'' in \emph{2023 IEEE 28th International Conference on Emerging Technologies and Factory Automation (ETFA)}.\hskip 1em plus 0.5em minus 0.4em\relax IEEE, 2023, pp. 1--8.

\bibitem{koziolek2024llm-retrieval}
H.~Koziolek, S.~Gr{\"u}ner, R.~Hark, V.~Ashiwal, S.~Linsbauer, and N.~Eskandani, ``Llm-based and retrieval-augmented control code generation,'' in \emph{Proceedings of the 1st International Workshop on Large Language Models for Code}, 2024, pp. 22--29.

\bibitem{koziolek2024llm-image}
H.~Koziolek and A.~Koziolek, ``Llm-based control code generation using image recognition,'' in \emph{Proceedings of the 1st International Workshop on Large Language Models for Code}, 2024, pp. 38--45.

\bibitem{kroening2014cbmc}
D.~Kroening and M.~Tautschnig, ``Cbmc--c bounded model checker: (competition contribution),'' in \emph{Tools and Algorithms for the Construction and Analysis of Systems: 20th International Conference, TACAS 2014, Held as Part of the European Joint Conferences on Theory and Practice of Software, ETAPS 2014, Grenoble, France, April 5-13, 2014. Proceedings 20}.\hskip 1em plus 0.5em minus 0.4em\relax Springer, 2014, pp. 389--391.

\bibitem{lee2024formal}
J.~Lee and K.~Bae, ``Formal semantics and analysis of multitask plc st programs with preemption,'' in \emph{International Symposium on Formal Methods}.\hskip 1em plus 0.5em minus 0.4em\relax Springer, 2024, pp. 425--442.

\bibitem{lee2022bounded}
J.~Lee, S.~Kim, and K.~Bae, ``Bounded model checking of plc st programs using rewriting modulo smt,'' in \emph{Proceedings of the 8th ACM SIGPLAN International Workshop on Formal Techniques for Safety-Critical Systems}, 2022, pp. 56--67.

\bibitem{li2023starcoder}
R.~Li, L.~B. Allal, Y.~Zi, N.~Muennighoff, D.~Kocetkov, C.~Mou, M.~Marone, C.~Akiki, J.~Li, J.~Chim \emph{et~al.}, ``Starcoder: may the source be with you!'' \emph{arXiv preprint arXiv:2305.06161}, 2023.

\bibitem{li2022competition}
Y.~Li, D.~Choi, J.~Chung, N.~Kushman, J.~Schrittwieser, R.~Leblond, T.~Eccles, J.~Keeling, F.~Gimeno, A.~Dal~Lago \emph{et~al.}, ``Competition-level code generation with alphacode,'' \emph{Science}, vol. 378, no. 6624, pp. 1092--1097, 2022.

\bibitem{liu2024large}
J.~Liu, K.~Wang, Y.~Chen, X.~Peng, Z.~Chen, L.~Zhang, and Y.~Lou, ``Large language model-based agents for software engineering: A survey,'' \emph{arXiv preprint arXiv:2409.02977}, 2024.

\bibitem{minsky1961steps}
M.~Minsky, ``Steps toward artificial intelligence,'' \emph{Proceedings of the IRE}, vol.~49, no.~1, pp. 8--30, 1961.

\bibitem{mordor_plc_market_2024}
\BIBentryALTinterwordspacing
{Mordor Intelligence}, ``Programmable logic controller (plc) market - share, size \& growth,'' 2024, accessed: 2024-10-08. [Online]. Available: \url{https://www.mordorintelligence.com/industry-reports/programmable-logic-controller-plc-market}
\BIBentrySTDinterwordspacing

\bibitem{nunez2024autosafecoder}
A.~Nunez, N.~T. Islam, S.~K. Jha, and P.~Najafirad, ``Autosafecoder: A multi-agent framework for securing llm code generation through static analysis and fuzz testing,'' \emph{arXiv preprint arXiv:2409.10737}, 2024.

\bibitem{ovatman2016overview}
T.~Ovatman, A.~Aral, D.~Polat, and A.~O. {\"U}nver, ``An overview of model checking practices on verification of plc software,'' \emph{Software \& Systems Modeling}, vol.~15, no.~4, pp. 937--960, 2016.

\bibitem{qian2024chatdev}
C.~Qian, W.~Liu, H.~Liu, N.~Chen, Y.~Dang, J.~Li, C.~Yang, W.~Chen, Y.~Su, X.~Cong \emph{et~al.}, ``Chatdev: Communicative agents for software development,'' in \emph{Proceedings of the 62nd Annual Meeting of the Association for Computational Linguistics (Volume 1: Long Papers)}, 2024, pp. 15\,174--15\,186.

\bibitem{chatdev}
\BIBentryALTinterwordspacing
C.~Qian, W.~Liu, H.~Liu, N.~Chen, Y.~Dang, J.~Li, C.~Yang, W.~Chen, Y.~Su, X.~Cong, J.~Xu, D.~Li, Z.~Liu, and M.~Sun, ``Chatdev: Communicative agents for software development,'' \emph{arXiv preprint arXiv:2307.07924}, 2023. [Online]. Available: \url{https://arxiv.org/abs/2307.07924}
\BIBentrySTDinterwordspacing

\bibitem{ribeiro2002reinforcement}
C.~Ribeiro, ``Reinforcement learning agents,'' \emph{Artificial intelligence review}, vol.~17, pp. 223--250, 2002.

\bibitem{roziere2023code}
B.~Roziere, J.~Gehring, F.~Gloeckle, S.~Sootla, I.~Gat, X.~E. Tan, Y.~Adi, J.~Liu, R.~Sauvestre, T.~Remez \emph{et~al.}, ``Code llama: Open foundation models for code,'' \emph{arXiv preprint arXiv:2308.12950}, 2023.

\bibitem{schreyer2000design}
M.~Schreyer and M.~M. Tseng, ``Design framework of plc-based control for reconfigurable manufacturing systems,'' in \emph{Proceedings of international conference on flexible automation and intelligent manufacturing (FAIM 2000)}, vol.~1, 2000, pp. 33--42.

\bibitem{shi2024automated}
J.~Shi, Y.~Chen, Q.~Li, Y.~Huang, Y.~Yang, and M.~Zhao, ``Automated test cases generator for iec 61131-3 structured text based dynamic symbolic execution,'' \emph{IEEE Transactions on Computers}, 2024.

\bibitem{siddiq2023lightweight}
M.~L. Siddiq, B.~Casey, and J.~Santos, ``A lightweight framework for high-quality code generation,'' \emph{arXiv preprint arXiv:2307.08220}, 2023.

\bibitem{singh2023taxonomy}
A.~Singh, ``Taxonomy of machine learning techniques in test case generation,'' in \emph{2023 7th International Conference on Intelligent Computing and Control Systems (ICICCS)}.\hskip 1em plus 0.5em minus 0.4em\relax IEEE, 2023, pp. 474--481.

\bibitem{tang2024collaborative}
D.~Tang, Z.~Chen, K.~Kim, Y.~Song, H.~Tian, S.~Ezzini, Y.~Huang, and J.~K. T.~F. Bissyande, ``Collaborative agents for software engineering,'' \emph{arXiv preprint arXiv:2402.02172}, 2024.

\bibitem{Technavio_plc_market_2023}
\BIBentryALTinterwordspacing
{Technavio}, ``Programmable logic controller (plc) market analysis apac, north america, europe, middle east and africa, south america - us, china, japan, germany, uk - size and forecast 2024-2028,'' 2024, accessed: 2024-10-08. [Online]. Available: \url{https://www.technavio.com/report/programmable-logic-controller-plc-market-industry-analysis}
\BIBentrySTDinterwordspacing

\bibitem{wang2023k}
K.~Wang, J.~Wang, C.~M. Poskitt, X.~Chen, J.~Sun, and P.~Cheng, ``K-st: A formal executable semantics of the structured text language for plcs,'' \emph{IEEE Transactions on Software Engineering}, 2023.

\bibitem{wang2024survey}
L.~Wang, C.~Ma, X.~Feng, Z.~Zhang, H.~Yang, J.~Zhang, Z.~Chen, J.~Tang, X.~Chen, Y.~Lin \emph{et~al.}, ``A survey on large language model based autonomous agents,'' \emph{Frontiers of Computer Science}, vol.~18, no.~6, p. 186345, 2024.

\bibitem{wang2021application}
M.~Wang, ``Application of plc technology in electrical engineering and automation control,'' in \emph{Application of Intelligent Systems in Multi-modal Information Analytics: Proceedings of the 2020 International Conference on Multi-model Information Analytics (MMIA2020), Volume 2}.\hskip 1em plus 0.5em minus 0.4em\relax Springer, 2021, pp. 131--135.

\bibitem{wu2023autogen}
Q.~Wu, G.~Bansal, J.~Zhang, Y.~Wu, B.~Li, E.~Zhu, L.~Jiang, X.~Zhang, S.~Zhang, J.~Liu, A.~H. Awadallah, R.~W. White, D.~Burger, and C.~Wang, ``Autogen: Enabling next-gen llm applications via multi-agent conversation framework,'' in \emph{COLM}, 2024.

\bibitem{xi2023rise}
Z.~Xi, W.~Chen, X.~Guo, W.~He, Y.~Ding, B.~Hong, M.~Zhang, J.~Wang, S.~Jin, E.~Zhou \emph{et~al.}, ``The rise and potential of large language model based agents: A survey,'' \emph{arXiv preprint arXiv:2309.07864}, 2023.

\bibitem{xia2023keep}
C.~S. Xia and L.~Zhang, ``Keep the conversation going: Fixing 162 out of 337 bugs for \$0.42 each using chatgpt,'' \emph{arXiv preprint arXiv:2304.00385}, 2023.

\bibitem{xu2024automated}
K.~Xu, G.~L. Zhang, X.~Yin, C.~Zhuo, U.~Schlichtmann, and B.~Li, ``Automated c/c++ program repair for high-level synthesis via large language models,'' in \emph{Proceedings of the 2024 ACM/IEEE International Symposium on Machine Learning for CAD}, 2024, pp. 1--9.

\bibitem{zhang2024codeagent}
K.~Zhang, J.~Li, G.~Li, X.~Shi, and Z.~Jin, ``Codeagent: Enhancing code generation with tool-integrated agent systems for real-world repo-level coding challenges,'' \emph{arXiv preprint arXiv:2401.07339}, 2024.

\bibitem{zhang2024survey}
Z.~Zhang, X.~Bo, C.~Ma, R.~Li, X.~Chen, Q.~Dai, J.~Zhu, Z.~Dong, and J.-R. Wen, ``A survey on the memory mechanism of large language model based agents,'' \emph{arXiv preprint arXiv:2404.13501}, 2024.

\end{thebibliography}


\end{document}